# Heat capacity and magnetocaloric effect in polycrystalline $Gd_{1-x}Sm_xMn_2Si_2$

Pramod Kumar, Niraj K. Singh, K. G. Suresh[a]

*Department of Physics, I. I. T .Bombay, Mumbai 400076, India*

A. K. Nigam and S. K. Malik

*Tata Institute of Fundamental Research, Homi Bhabha Road, Mumbai 400005, India*

**Abstract**

We report the magnetocaloric effect in terms of isothermal magnetic entropy change as well as adiabatic temperature change, calculated using the heat capacity data. Using the zero field heat capacity data, the magnetic contribution to the heat capacity has been estimated. The variations in the magnetocaloric behavior have been explained on the basis of the magnetic structure of these compounds. The refrigerant capacities have also been calculated for these compounds.

## I. INTRODUCTION

Intermetallic compounds formed between rare earths (R) and transition metals (TM) have attracted considerable attention owing to their potential for various applications[1-3]. Recently, there has been a vigorous activity in the development of magnetocaloric materials, which are the active materials in magnetic refrigerators[4-7]. The magnetocaloric effect (MCE) manifests as the isothermal magnetic entropy change or the adiabatic temperature change of a magnetic material when exposed to a varying magnetic field. This effect is an intrinsic property of a magnetic material which occurs due to the change in the degree of alignment of the magnetic moments of the system under the influence of an applied magnetic field. The concept of magnetic refrigeration, which is based on MCE, has attracted a great deal of attention from several researchers and has triggered an intensive search for materials with giant MCE. Generally, due to their large magnetic moments, heavy rare earth elements and their compounds are considered as the best-suited materials for achieving large MCE. After the observation[8] of a giant MCE in $Gd_5Si_2Ge_2$, much focus has been given on the compounds showing field-induced magnetic phase transitions and/or structural transitions. But the magnetocaloric studies in $MnFeP_{1-x}As_x$ system[9] has shown that large MCE is not just restricted to the compounds with large magnetic moments but depends strongly on the type of magnetic/structural phase transitions as well.

Among the R-TM intermetallics, $RMn_2Si_2$(R=Gd, Sm) have attracted a lot of attention due to their interesting magnetic properties [10-13]. These compounds, in general, crystallize in the tetragonal ThCr$_2$Si$_2$-type structure. The magnetic properties of this system are quiet interesting because of the differences in the ordering temperatures of R and Mn sublattices. Recently, as part of our investigations on the magnetic and magnetocaloric properties, we have reported the MCE associated with the rare earth magnetic ordering in Gd$_{1-x}$Sm$_x$Mn$_2$Si$_2$ compounds[14]. It was found that the rare earth ordering temperature of these compounds ($T_C^R$) lies in the range of 37-65 K. The MCE was calculated using the magnetization isotherms close to $T_C^R$. In this paper, we report the magnetocaloric effect in terms of both isothermal entropy change (-$\Delta S_M$) and adiabatic temperature change ($\Delta T_{ad}$) calculated using the heat capacity data collected over a wide temperature range, both in zero field and in presence of magnetic field. The magnetic contribution to the total heat capacity has been obtained.

## II. EXPERIMENTAL DETAILS

The samples preparation of Gd$_{1-x}$Sm$_x$Mn$_2$Si$_2$ [x=0, 0.4, 0.6 and 1] compounds reported earlier[14]. The heat capacity (C) was measured using the relaxation method in the temperature (T) range of 1.5–280 K and in fields (H) up to 50 kOe, with the help of a physical property measurement system (PPMS, Quantum Design).

## III. RESULTS AND DISCUSSION

Fig.1 shows the temperature variation of heat capacity of Gd$_{1-x}$Sm$_x$Mn$_2$Si$_2$ compounds under zero field. The λ-like peak observed at temperatures close to $T_C^R$ in zero field indicates the second order nature of the transition. The inset of Fig.1 shows the temperature variation of heat capacity of SmMn$_2$Si$_2$ in zero field as well as in 50 kOe. As can be seen form the inset, with increase in the field, the peak at $T_C^R$ becomes broader and shifts to higher temperatures.

In general, the heat capacity of metallic magnetic systems can be written as $C_{tot}=C_{ph}+C_{el}+C_M$, where $C_{el}$, $C_{ph}$ and $C_M$ are the electronic, lattice and magnetic contributions respectively. In order to separate $C_M$ from $C_{tot}$, the first two terms have to be evaluated. This has been done by taking into account the iso-structural nonmagnetic counterparts namely LaFe$_2$Si$_2$, whose heat capacity can be written as

$$C_{tot} = \gamma T + 9NR(T/\theta_D)^3 \int_0^{\theta_D/T} \frac{x^4 e^x}{(e^x-1)^2}dx \qquad (1)$$

Where the first term represents the electronic contribution and the second term corresponds to the phonon contribution. N is the number of atoms per formula unit (N=5 in this case), R is the molar gas constant, γ is the electronic coefficient and $\theta_D$ is the Debye temperature. It has been reported that the heat capacity of LaFe$_2$Si$_2$ follows the above expression satisfactorily with an effective $\theta_D$=280K and γ=22.7mJ/mol K, respectively. By using the relation $\theta_D \propto M^{-\frac{1}{2}}$, we have calculated the Debye temperature of the present series of compounds. The sum of the nonmagnetic contributions to the heat capacity has been calculated and the representative plot for Gd$_{0.4}$Sm$_{0.6}$Mn$_2$Si$_2$ is shown in Fig.2 (red solid lines). The temperature dependence of the magnetic contribution (filled squares) calculated for this compound is shown by filled squares. A similar analysis has been done for other compounds as well.

Fig.3 shows the variation of magnetic entropy ($S_M$) with temperature for all the compounds. It may be noticed from this figure that the curves do not saturate even temperatures well above $T_C^R$. At $T_C^R$, the values of magnetic entropy are 13.1, 16, 13.4 and 9.1 J/mol K for the compounds with x=0, 0.4, 0.6 and 1, respectively. It is of interest to note that $R \ln (2J+1)$ value for $Gd^{3+}$ ion (J=7/2) and $Sm^{3+}$ ion (J= 5/2) are 17.2 J/mol K and 16 J/mol K, respectively. The magnetic entropy below $T_C^R$ originates mainly due to the rare earth sublattice. Almost the full entropy associated with the rare earth moment is released at $T_C^R$, above which the entropy of the rare earth sublattice should be nearly temperature independent. On the other hand, the Mn sublattice contributes to $S_M$ mainly at temperatures above $T_C^R$, because of the fact that the Neel temperature of the Mn sublattice is in the range of 380-480 K in these compounds. Therefore, the curvature seen in the $S_M$ vs. T curves above $T_C^R$ is due to the Mn sublattice, whose antiferromagnetic ordering decreases with temperature.

The magnetocaloric effect of $Gd_{1-x}Sm_xMn_2Si_2$ compounds calculated from the C-H-T data has been calculated using the methods reported earlier[15-16]. Figs. 4a&b show the temperature variation of isothermal magnetic entropy and adiabatic temperature change for a field of 50kOe, for all compounds. $\Delta S_M$ values calculated from the C-H-T data were found to be in agreement with those obtained from the M-H-T data, reported earlier[14]. $\Delta S_M$ vs. T plot shows a maximum near $T_C^R$ with the value of 3.9 J/kg K and 4.2 J/kg K for x=0 and 1 respectively. The maximum value of $\Delta S_M$ ($\Delta S_M^{max}$) is found to decrease with Sm concentration, but in $SmMn_2Si_2$ the value of MCE is slightly larger than GdMn$_2$Si$_2$. Moreover, the width of the MCE peak is considerably smaller in SmMn$_2$Si$_2$, as compared to the other compounds. In fact the peaks in the compounds with x=0, 0.4 and 0.6 show some structure, possibly indicating the presence of some other contributions at low temperatures. As reported earlier[14], R-Mn interaction is antiferromagnetic for heavy rare earths and ferromagnetic for light rare earths. Therefore, Gd-Mn interaction is ferrimagnetic and Sm-Mn interaction is ferromagnetic in these compounds. Due to this reason, the magnetic state in the intermediate concentrations (i.e. x=0.4 and 0.6) is complex and this may be the reason for the anomalous nature of the MCE plots of these compounds. The anomalous shape of the MCE plots in the present case may be due to the following reason. Based on the magnetoresistance data, Wada et al.[17] have recently suggested a new type of magnetic state for GdMn$_2$X$_2$ compound at temperatures below $T_C^R$. Since Gd-Mn coupling is ferrimagnetic, with increase in applied field, the net field acting on the Mn moments decreases and as a result, the Mn sublattice tends to become antiferromagnetically coupled. Consequently, a canted magnetic structure develops at temperatures below $T_C^R$ in presence of an applied field. Fujiwara et al.[18] have also reported a canted magnetic structure for RMn$_2$X$_2$ compounds. We feel that this canted structure is responsible for the unusually large magnetic entropy change at temperatures much below $T_C^R$, resulting in broad MCE peaks in the compounds with x=0, 0.4 and 0.6. On the other hand, in SmMn$_2$Si$_2$, Sm-Mn coupling is ferromagnetic and hence there is no such possibility of a canted structure. This may be the reason for the sharper MCE peak in this compounds compared to that of other compounds of this series. As is evident from Fig.4b, the temperature variation of adiabatic temperature change is similar to that of the entropy change. $\Delta S_M$ can be also explained qualitatively on the basis of the molecular field model as applied to the two sublattices of R and Mn.

The refrigerant capacity (RC) have been calculated for both the compounds using the formula

$$q = \int_{T_1}^{T_2} \Delta S_M(T) dT \quad (2)$$

Were $T_1$ and $T_2$ are the temperatures of the hot and cold reservoirs, respectively. In order to have a meaningful comparison with other materials, the RC for the present compounds has been calculated in units of J/kg. The full-width at half-maximum (FWHM) of the MCE peaks obtained for x=0, 0.4, 0.6 and 1 has been found to be 56, 33.8, 40.9 and 41.3 K, respectively. The RC values were obtained as 220, 118, 63 and 83 J/kg for x=0, 0.4, 0.6 and 1, respectively. The temperatures corresponding to the FWHM points have been taken as the limits of integration in eqn. 2. The RC observed in the present case compare well with those of many potential materials[19,20] whose magnetic transitions are in the same range as in the present case. Table 1 shows the magnetocaloric properties along with the rare earth ordering temperatures in these compounds.

**Acknowledgement**

One of the authors (KGS) thanks ISRO, Govt. of India for supporting this work through a sponsored research grant.

Table 1 Rare earth ordering temperature, isothermal magnetic entropy change, adiabatic temperature change and the refrigerant capacity in $Gd_{1-x}Sm_xMn_2Si_2$ compounds.

| x | $T_C^R$ (K) | $-(\Delta S_M)^{max}$ (J/kg K) | $\Delta T_{ad}$ (K) | RC (J/kg) |
|---|---|---|---|---|
| 0 | 63 | 3.9 | 1.5 | 220 |
| 0.4 | 53 | 3.4 | 1.8 | 118 |
| 0.6 | 48 | 1.4 | 0.76 | 63 |
| 1.0 | 37 | 4.2 | 2.4 | 83 |

List of Figures:

Fig.1 Temperature variation of heat capacity of $Gd_{1-x}Sm_xMn_2Si_2$ compounds in zero field. The inset shows the variation in $SmMn_2Si_2$ under zero field and 50 kOe.

Fig.2 Temperature variation of heat capacity of $Gd_{0.4}Sm_{0.6}Mn_2Si_2$ in zero field (open circles). The solid line represents the theoretically calculated electronic and phonon contributions ($C_{Th}$). The solid squares represent the magnetic contribution ($C_M$).

Fig.3 Temperature variation of magnetic entropy ($S_M$) of $Gd_{1-x}Sm_xMn_2Si_2$ compounds.

Fig.4a-b Temperature variation of (a) isothermal magnetic entropy change ($-\Delta S_M$) and (b) adiabatic temperature change ($\Delta T_{ad}$) in $Gd_{1-x}Sm_xMn_2Si_2$ compounds for a field change of 50kO.

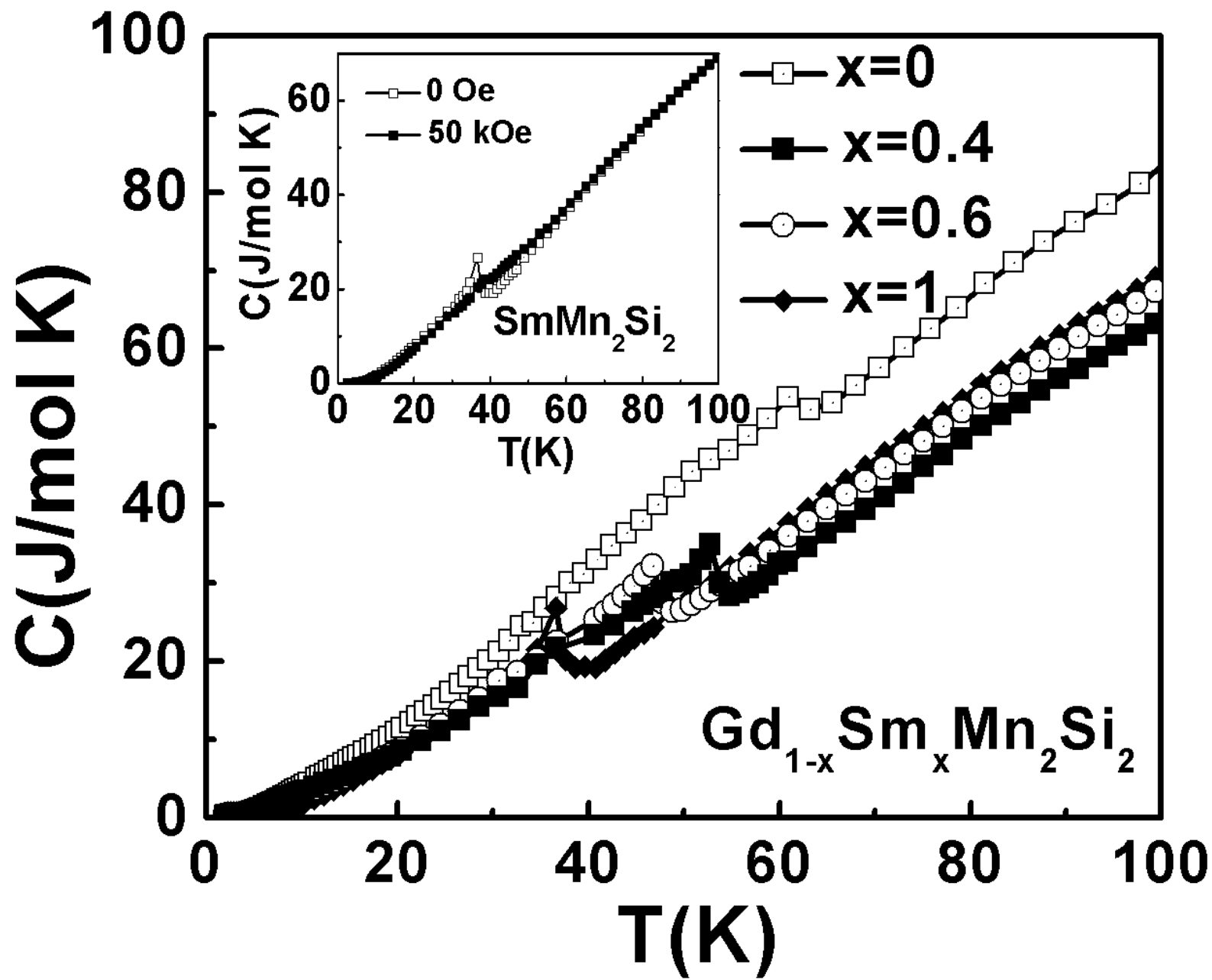

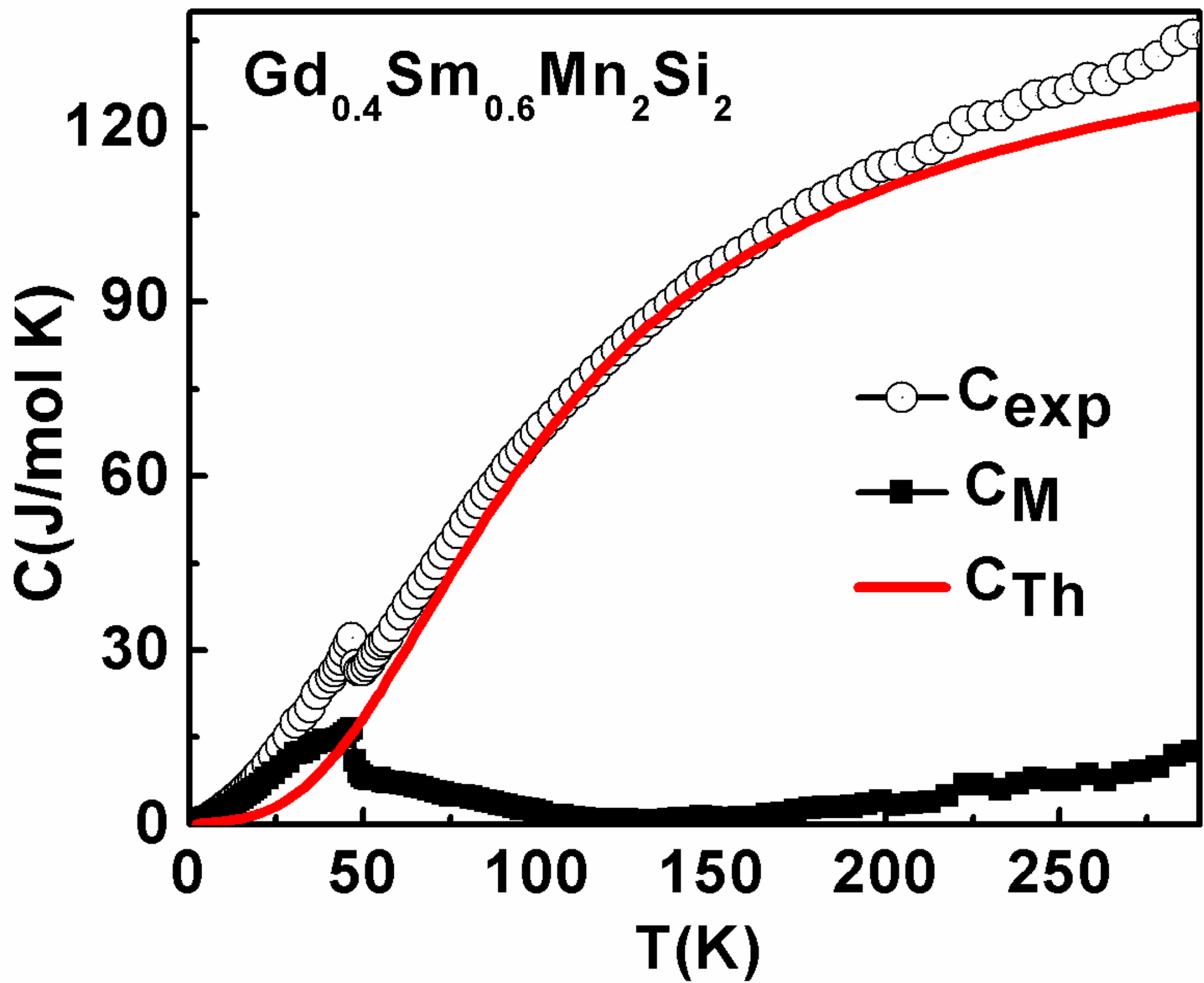

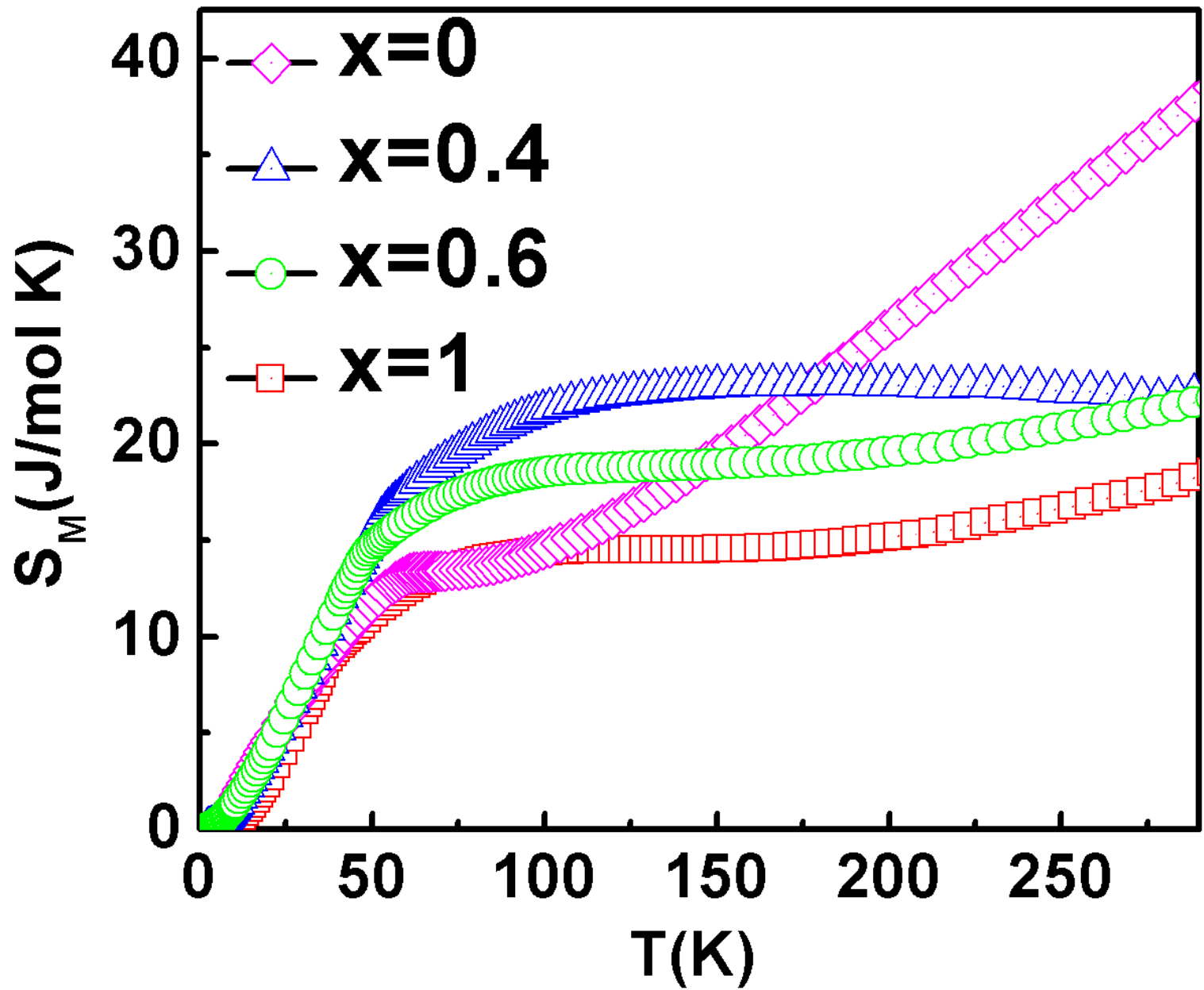

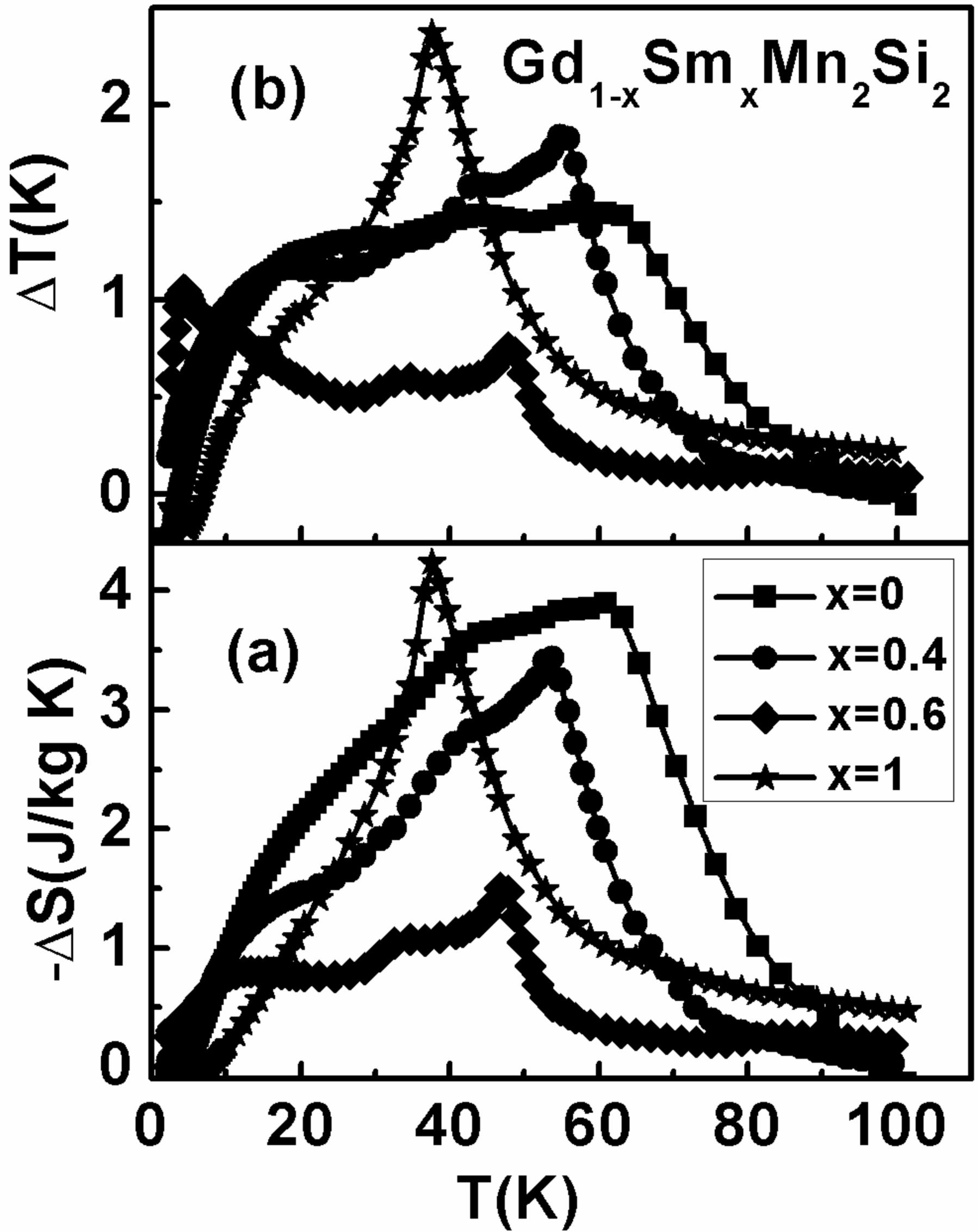